\documentclass{article}
\usepackage{spconf,amsmath,epsfig}
\usepackage{hyperref}
\usepackage{xcolor}

\usepackage{graphicx}
\usepackage{amssymb}
\graphicspath{ {images/} }

\usepackage{subcaption}
\usepackage{flushend}

\let\OLDthebibliography\thebibliography
\renewcommand\thebibliography[1]{
  \OLDthebibliography{#1}
  \setlength{\parskip}{0pt}
  \setlength{\itemsep}{0pt plus 0.3ex}
}

\usepackage{fancyhdr}
\fancypagestyle{firstpage}{
  \fancyhf{}
  \fancyhead[C]{\vspace{-2.2cm} \small Accepted for presentation at the Second IEEE ICME Workshop on Artificial Intelligence in Sports (AI-Sports), Shenzhen, China, July 2021.\\© 2021 IEEE. Personal use of this material is permitted. Permission from IEEE must be obtained for all other uses, in any current or future media, including reprinting/republishing this material for advertising or promotional purposes, creating new collective works, for resale or redistribution to servers or lists, or reuse of any copyrighted component of this work in other works. \vspace{0.3cm}} 
}

\begin{document}\sloppy

\def\x{{\mathbf x}}
\def\L{{\cal L}}


\title{Swimmer Stroke Rate Estimation From Overhead Race Video}
%
\name{Timothy Woinoski and Ivan V. Baji\'{c}\thanks{978-1-6654-3864-3/21/\$31.00 ©2021 IEEE}}
\address{School of Engineering Science, Simon Fraser University, Canada}

\maketitle

\begin{abstract}
In this work, we propose a swimming analytics system for automatically determining swimmer stroke rates from overhead race video (ORV). General ORV is defined as any footage of swimmers in competition, taken for the purposes of viewing or analysis. Examples of this are footage from live streams, broadcasts, or specialized camera equipment, with or without camera motion. These are the most typical forms of swimming competition footage. We detail how to create a system that will automatically collect swimmer stroke rates in any competition, given the video of the competition of interest. With this information, better systems can be created and additions to our analytics system can be proposed to automatically extract other swimming metrics of interest.
\end{abstract}
\thispagestyle{firstpage}

\section{Introduction} \label{sec:intro}
In this work, we propose a swimming analytics system for automatically determining swimmer stroke rates from overhead race video (ORV). We define General ORV as any footage of swimmers in competition, taken above the water level for the purposes of viewing or analysis. Examples of this include footage from live streams, broadcasts, or specialized camera equipment, with or without camera motion. Much of the data requested by swimming coaches, such as stroke rates for swimmer analytics, can be manually extracted from ORV. We aim to automatically extract this data as extraction can be very time-consuming and error-prone for humans. 

No system that completely automates the analysis of general swimming ORV currently exists. However, a system that takes specialized static ORV of an entire pool has been proposed in~\cite{hall2020detection}. Our proposed system differs in that the camera is allowed to move and capture only a portion of the pool, which covers most typical ORV scenarios from swimming broadcasts and live streams. In particular, our system takes advantage of the developments in deep models for computer vision to allow greater flexibility regarding the input video that can be used for extracting relevant swimming analytics. 

The paper is organized as follows. In Section~\ref{sec:currentWork}, relevant prior work is reviewed. Section~\ref{sec:dataset} describes the dataset we have developed for training and testing, which we are releasing with the paper. Section~\ref{sec:proposedMeth} describes the proposed system, including its modules, detection, tracking, and stroke rate estimation. Experiments are described in Section~\ref{sec:experiments}, followed by a conclusion in Section~\ref{sec:conclusion}.

\section{Related Work} \label{sec:currentWork}
Methods of computer vision and human pose estimation have been considered for swimming analytics for years. Initially, work was conducted to collect analytics in highly controlled environments~\cite{zecha}. For example, underwater footage was analyzed from a swimming channel. This footage was very different from the general ORV, and the setup could not easily translate to general race swimming, but the approach showed that automated swimming analytics had promise. Later, ORV was proposed as a source of data for automated collection of swimming analytics~\cite{hall2020detection,sha,sha_understanding,victor,victor_new}. 

In~\cite{sha,sha_understanding}, a calibrated lab-like setup was proposed for automated swimming analytics from ORV. The authors used a complex assortment of tracking algorithms, human proposed regions of interest, and a multitude of hand-crafted computer vision methods in order to analyze swimmers. While excellent results were obtained, the methods were tailored to the one pool their work was dedicated to. A highly complex tuning process would have to be repeated for each new pool to be analyzed. More recently,~\cite{victor,victor_new} presented a novel method for extracting the stroke rate of a swimmer given a video of a single swimmer. As ORV does not usually track a single swimmer for the entirety of a race, this approach cannot be directly applied to general ORV. 

The most recent published work on this topic~\cite{hall2020detection} presents a system called DeepDASH, in which all swimmers' strokes could be automatically extracted from static, wide-angle high definition ORV that captures the entire pool at all times. Our proposed approach allows for a similar set of analytics to be extracted, but from a more general broadcast-style ORV, which is much more common and widely available.  

\section{Swimmer Analytics Dataset} \label{sec:dataset}
As part of this research, in addition to the methods described in the paper, a swimming analytics dataset, with data for swimmer detection and stroke rate estimation, has been provided alongside a custom annotation software.\footnote{\url{https://github.com/tjwoinosk/swim\_annotator}} The dataset consists of YouTube ORV clips and manually generated labels that can be used to train machine learning models.

\subsection{Swimmer Detection Dataset} \label{sec:detDataset}
To our knowledge, no public dataset currently exists for swimmer detection in the general broadcast-style ORV. The authors of~\cite{hall2020detection} created their own dataset annotating the heads of swimmers from nine different venues, using video from a static camera. In contrast, our dataset consists of 35 broadcast-style ORV videos containing three camera views with over 50\% non-static footage, all taken from a single pool. In addition, the annotations outline the entire body of a swimmer rather than just the head. Such data can be used to train popular object detectors~\cite{girshick2014rich, liu2016ssd, lin2017focal, Redmon} for detecting swimmers in a racing environment. 

The ground truth was collected from videos following the definitions in~\cite{woinoski2020towards} using custom software mentioned. We utilize videos of swimmers collected from a competition posted on Swim USA's YouTube page~\cite{swim_usa}. This competition is from the 2019 TYR Pro Swim Series, and we refer to it as the Bloomington competition. The videos had resolutions of 720$\times$1280 and 1080$\times$1920, respectively, and were taken at 30 frames per second (fps). In addition, we also included the lane number of the swimmer in the venue. We annotated every third frame of the video except for videos capturing the diving class, in which every frame was annotated. This resulted in roughly 3,000 annotated frames. Swimmers' poses were annotated in six classes (``on-blocks'', diving, swimming, underwater, turning, and finishing), and the proportion of each class in the dataset is shown in Table~\ref{tab:collected_data}. This data was used as the training set. The test set consisted of roughly 250 frames from a different race, which was not used in the training. 

\begin{table}
    \centering
    \begin{tabular}{l|r|r}
        Class & \# Annotations & \% of Total\\
        \hline \hline
        ``On-blocks'' & 2,344 \hspace{0.45cm} & 10\% \hspace{0.5cm} \\
        Diving & 1,124 \hspace{0.45cm} & 5\%\hspace{0.6cm} \\
        Swimming & 13,009 \hspace{0.45cm} & 53\%\hspace{0.6cm} \\
        Underwater & 2,997 \hspace{0.45cm} & 12\%\hspace{0.6cm} \\
        Turning & 1,558 \hspace{0.45cm} & 6\%\hspace{0.6cm} \\
        Finishing & 3,534 \hspace{0.45cm} & 14\%\hspace{0.6cm} \\
        \hline
        Total & 24,566 \hspace{0.45cm} & 100\%\hspace{0.6cm} \\
    \end{tabular}
    \caption{The amount of collected data for each class.}
    \label{tab:collected_data}
\end{table}


\subsection{Stroke Rate Estimation Dataset} \label{sec:stroke_recognition_data}
The stroke rate estimation dataset was also created from the Bloomington competition, 
but the races were chosen to be different in terms of event and swimmers from those used for the detection dataset described above. First, each race video was split into sub-videos, each sub-video containing only one swimmer, using our tracking system described in Section~\ref{sec:proposedMeth}. Then, for each sub-video, we assigned a stroke value, called \textit{s-value}, for every frame in that video. This was done by fitting a sinusoidal curve with range $[0,1]$ to every stroke cycle in any given sub-video. An s-value of $1$ signifies the top of a stroke, 
while an s-value of $0$ indicates the bottom of a stroke. 
In addition to this, for every frame, a flag is given detailing if the swimmer in the frame was in the swimming class. If the swimmer was not swimming (e.g., if turning) their s-value was set to $0.5$. 
With this procedure, data from 346 sub-videos containing roughly 188,000 s-values was created.

\section{Methods} \label{sec:proposedMeth}
\begin{figure}
\centering
\includegraphics[width=\linewidth]{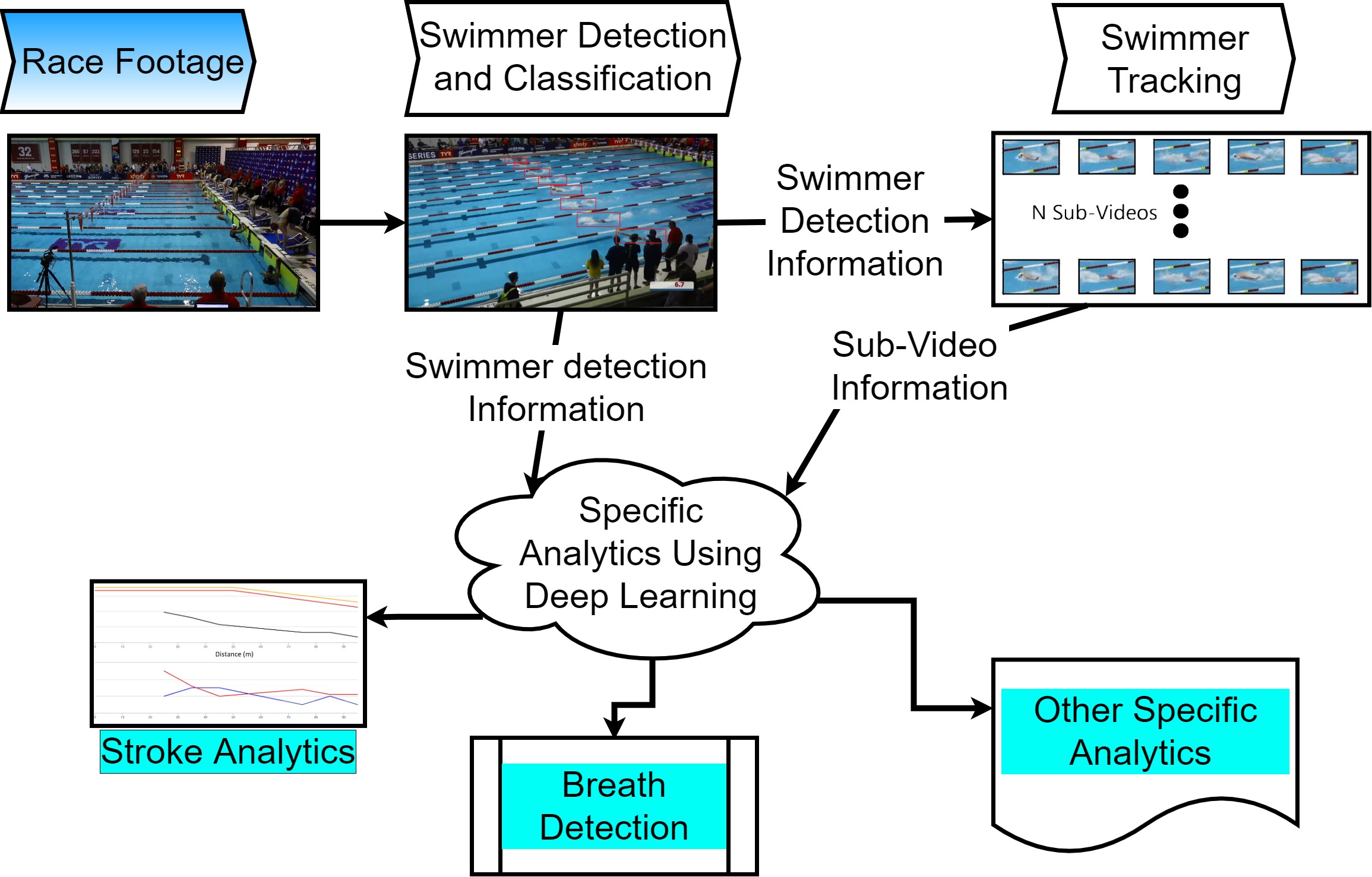}
\caption{The envisioned analytics system.}
\label{fig:order_of_race}
\end{figure}

The envisioned swimming analytics system is shown in Fig.~\ref{fig:order_of_race}. In this section, we describe swimmer detection, classification, tracking, and stroke rate estimation. Other analytics systems shown in Fig.~\ref{fig:order_of_race}, such as breath detection, can be built on the foundation presented here, but are left for future work. 

\subsection{Swimmer Detection} \label{sec:detection}
We selected the YOLOv3~\cite{yolov3,tiny_yolo} model architecture for swimmer detection and classification. YOLOv3 was chosen for its inference speed while offering competitive accuracy. Both a YOLOv3-416-tiny, with a Darknet15 backbone, and the YOLOv3-416, with a Darknet53 backbone, were used for swimmer detection. Transfer learning from the COCO-dataset pre-trained weights was used to train detectors on the swimmer detection dataset described in Section~\ref{sec:detDataset}. The hyperparameters for training, such as the number of epochs, learning rate, burn-in, batch size, and model anchors were kept at default values. 
Each model was trained to 12,000 epochs, as suggested in~\cite{AlexeyAB_darknet}. The trained models output a bounding box for each detected swimmer, along with the class (Table~\ref{tab:collected_data}) with the highest confidence value.



\begin{figure*}[t]
    \centering
    \begin{subfigure}[b]{0.43\textwidth}
        \centering
        \includegraphics[width=\textwidth]{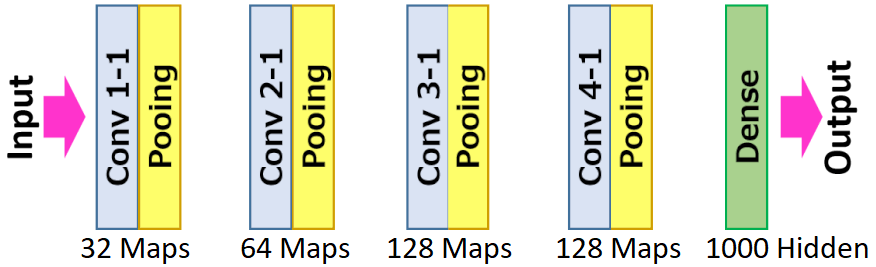}
        \caption{The ``small'' model for stroke recognition}
        \label{fig:small_arch}
    \end{subfigure}
    \begin{subfigure}[b]{0.54\textwidth}
        \centering
        \includegraphics[width=\textwidth]{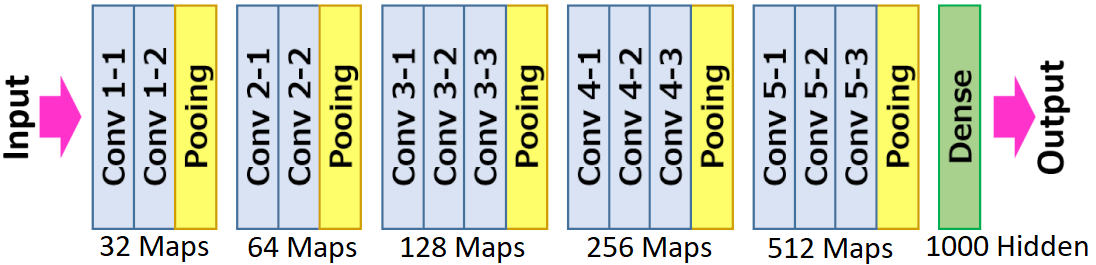}
        \caption{The VGG16-based model for stroke recognition}
        \label{fig:vgg_arch}
    \end{subfigure}
    \caption{Two of the models used for stroke recognition in the proposed system.}
    \label{fig:new_arch}
\end{figure*}

\subsection{Swimmer Tracking} \label{sec:tracking}
For swimmer tracking, we adopted the Simple Online and Realtime Tracking (SORT)~\cite{bewley2016simple} approach, which consists of detection, tracking, and data association. In our case, swimmer detections are provided by the re-trained YOLOv3, as described above; tracking is accomplished via Kalman filtering~\cite{Sayed_Adaptive_Filters_2008}, and data association using the Hungarian Algorithm~\cite{kuhn1955hungarian}. The Kalman filter uses a linear constant-velocity model. The state vector at time $k$ is given by~\cite{bewley2016simple}:
\begin{equation}\label{equ:state} 
    \mathbf{x}_k = [u,v,s,r,\dot{u},\dot{v},\dot{s}]^\top,
\end{equation}
where $u$ and $v$ represent the horizontal and vertical coordinates of the centroid of the swimmer's bounding box. For a bounding box of width $w$ and height $h$, $s$ and $r$ are defined as $s=w\cdot h$ and $r=w/h$. Finally, $\dot{u}$, $\dot{v}$, $\dot{s}$ are the time derivatives of $u$, $v$, and $s$. Note that the time derivative of $r$ is not part of the state vector, because it is assumed that the bounding box's aspect ratio $w/h$ does not change~\cite{bewley2016simple}. The prediction step consists of state prediction and error covariance prediction:
\begin{align}
    \widehat{\mathbf{x}}_k^{-} & = \mathbf{A}\widehat{\mathbf{x}}_{k-1}, \label{eq:Kalman_prediction}\\
    \mathbf{P}_k^{-} & = \mathbf{A}\mathbf{P}_{k-1}\mathbf{A}^\top + \mathbf{Q},
\end{align}
where $\mathbf{A} \in \mathbb{R}^{7\times7}$ is a matrix with all zeros except ones on the main diagonal and the sixth diagonal (so that $\widehat{u}_k^{-}=\widehat{u}_{k-1}+\widehat{\dot{u}}_{k-1}$, etc.), $\mathbf{P}_k \in \mathbb{R}^{7\times7}$ is the error covariance matrix at step $k$, and $\mathbf{Q} \in \mathbb{R}^{7\times7}$ is the process noise convariance matrix. The Kalman gain matrix at time $k$ is computed as
\begin{equation}
    \mathbf{K}_k = \mathbf{P}_k^{-}(\mathbf{P}_k^{-} + \mathbf{R})^{-1},
\end{equation}
where $\mathbf{R} \in \mathbb{R}^{7\times7}$ is the measurement noise covariance matrix. Finally, the update step consists of updating the error covariance matrix and the state estimate using the detected swimmers at time $k$:
\begin{align}
    \mathbf{P}_k &= (\mathbf{I}-\mathbf{K}_k)\mathbf{P}_k^{-},\\
    \widehat{\mathbf{x}}_k &= \widehat{\mathbf{x}}_k^{-} + \mathbf{K}_k(\mathbf{z}_k - \widehat{\mathbf{x}}_k^{-}),
\end{align}
where $\mathbf{z}_k$ corresponds to the detected swimmer at time $k$. Associations between detected bounding boxes produced by YOLOv3 and predicted boxes produced in~(\ref{eq:Kalman_prediction}) are made using the Hungarian Algorithm~\cite{kuhn1955hungarian}. For initialization, we used $\mathbf{P}_0=\mathbf{Q}=10^{-2}\mathbf{I}$ and $\mathbf{R}=10^{-1}\mathbf{I}$.

Using the tracked bounding boxes for each swimmer, a sub-video corresponding to each swimmer in the ORV can be created. Then these sub-videos are input to the stroke rate estimation module, described in the next section.

\subsection{Stroke Rate Estimation} \label{sec:action}
Stroke rate estimation is performed by feeding the frames of a sub-video tracking a single swimmer into a convolutional neural network (CNN), whose task is to predict the s-value (Section~\ref{sec:stroke_recognition_data}) of a stroke in the input frame. 
The produced sequence of s-values is smoothed using a low-pass Butterworth filter of order 8 with the cut-off ($-3$dB) frequency of $3$Hz, which corresponds to $180$ half-strokes per minute. The rationale is that no swimmer can achieve a higher stroke rate. The smoothed signal is then transformed into a square wave function with range $[0,1]$ by thresholding at the mean value of the smoothed output signal. Using this square wave, the mean position in time of each sequential string of $1$'s was assigned as the predicted position of the top of a stroke (s-value of $1$). 

Three CNN models for predicting s-values were explored in the experiments. One of them, which is referred to as ``Victor'', was proposed in~\cite{victor} and contains 16.9 million parameters. The other two are shown in Fig.~\ref{fig:new_arch}. The model shown in Fig.~\ref{fig:small_arch} is referred to as ``small''. It consists of a sequence of convolutional and max-pooling layers, followed by a 1000-unit dense layer, a single output unit with a sigmoid activation. Overall, it contains 1.0 million parameters. All convolutional layers consist of $3\times3$ filters, with a stride of~1, and Rectified Linear Unit (ReLU) activation. The number of filters in each layer is shown in the figure (referred to as ``Maps''). Max-pooling operates on windows of size $2\times2$ with a stride of 2.

The model shown in Fig.~\ref{fig:vgg_arch} is based on the VGG16 architecture and uses the pre-trained VGG16 backbone. It consists of convolution-pooling blocks, where each block is built up of two or three convolutional layers followed by a max-pooling layer. Overall, it contains 16.7 million parameters. As in the ``small'' model, convolutional layers consist of $3\times3$ filters, with a stride of 1, and ReLU activation. The number of filters in each layer is shown in the figure. Max-pooling operates on windows of size $2\times2$ with a stride of 2. The weights of the first four layers of the backbone were frozen during training.


\subsection{Training}
Training was conducted in Keras~\cite{chollet2015Keras} using the RMSprop optimizer with a learning rate of $10^{-4}$. 
The models were trained to output the s-value (Section~\ref{sec:stroke_recognition_data}) for each input frame. The loss function was the mean absolute error (MAE) between the model output and the ground-truth s-values. Training image pixel values were first scaled to values in the range $[1/255, 1]$. Training images were augmented by rotations, width/height shifts, shears, zooms, and horizontal flips, and the models were trained for 
35 epochs. 

\section{Experiments} \label{sec:experiments}

\subsection{Swimmer Detection and Classification} \label{sec:detExp}
An examination of the dataset, where every third frame out of 30fps videos was annotated, suggests that there is a lot of redundancy in the data. We, therefore, wanted to examine the behavior of detection models, developed based on YOLOv3 and YOLOv3-tiny architectures, as a function of the amount of data they were trained on. Specifically, we randomly selected subsets of various sizes from the training set, trained models on them, and then tested them on the test set to assess their performance.  

Five models with YOLOv3-416 architecture and five models with YOLOv3-tiny-416 architecture were trained on the following eleven training subset sizes: 1\%, 2\%, 5\%, 10\%, 15\%, 20\%, 25\%, 30\%, 50\%, 75\% ,and 100\% of the training set. In total, 110 different models were trained and tested, 55 for each of the YOLOv3-416 and YOLOv3-tiny-416 model architectures. Each of the models had the same training parameters and hyper-parameters; the only way they differed was in the (amount of) data they were trained on. The mean average precision (AP) with Intersection-over-Union (IoU) threshold of 25\% (AP-25) of each model was evaluated against the test set to examine the resulting accuracy and variability as a function of the training set size. 
We report the average precision (AP-25) on the ``swimming'' class, which contains detections of swimmers while swimming, and on ``not swimming'', which is the mean AP of the other five classes shown in Table~\ref{tab:collected_data}. Finally, the mean AP (mAP) of all classes is also reported.

\begin{figure}[t!]
    \centering
    \begin{subfigure}[b]{0.8\linewidth}
        \centering
        \includegraphics[width=\linewidth]{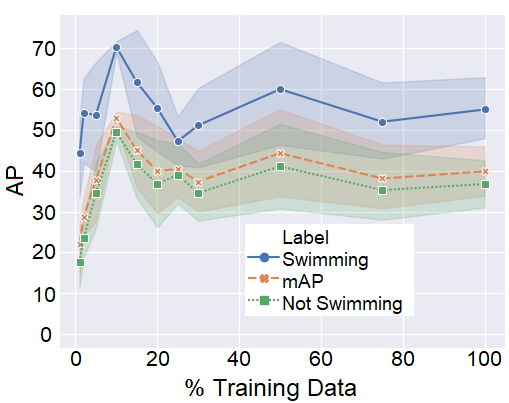}
        \caption{AP-25 of YOLOv3-416-tiny models}
    \end{subfigure}
    \hfill
    \begin{subfigure}[b]{0.8\linewidth}
        \centering
        \includegraphics[width=\linewidth]{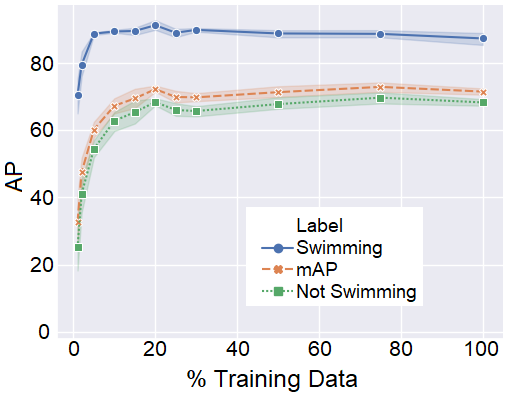}
        \caption{AP-25 of YOLOv3-416 models}
    \end{subfigure}
    \caption{Detection accuracy vs. percentage of the training set}
    \label{fig:res_plot}
\end{figure}

The results are shown in Fig.~\ref{fig:res_plot}. Since five models were trained for each training (sub)set size, we show their average AP-25 scores as well as the 95\% confidence interval around the average, in the form of a shaded band. The horizontal axis is labeled ``\% Training Data'' and means the percentage of the training set used for training the particular set of models. 
The results indicate that the model with the YOLOv3-416 architecture (Fig.~\ref{fig:res_plot}b) is more accurate than the smaller model based on the YOLOv3-416-tiny architecture (Fig.~\ref{fig:res_plot}a). In both cases, the ``swimming'' class is detected more accurately than ``not swimming'' classes, presumably because ``swimming'' class is the most represented in the dataset (Table~\ref{tab:collected_data}) and the model was exposed to more samples of this class than any other. Another conclusion from these results is that 10-20\% of the training data seems sufficient to reach a reasonable detection accuracy, because of the redundancy in the data. A recommendation for future swimming video annotation is that for the less-represented classes such as ``diving'' and ``turning'', each frame may be necessary, but for the ``swimming'' class, it is likely sufficient to annotate only one in 15 frames (i.e., 2 frames per second). Based on these results, we used a model with YOLOv3-416 architecture trained on 20\% of the training data for subsequent experiments.

\begin{figure*}[t]
    \centering
    \begin{subfigure}[b]{.32\textwidth}
        \centering
        \includegraphics[trim=0 0 0 200, clip, width=\textwidth]{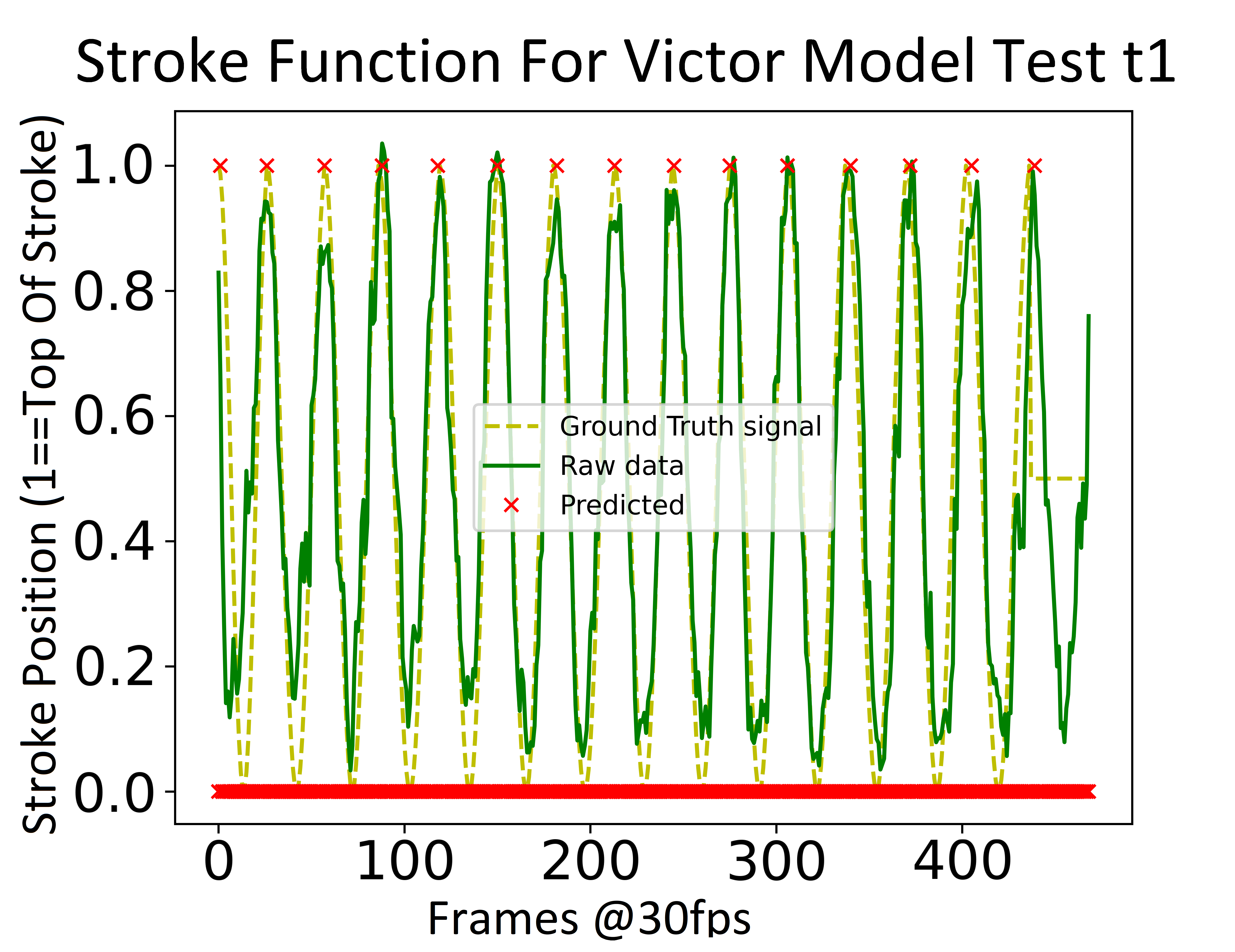}
        \caption{Victor model}
        \label{fig:t1_victor}
    \end{subfigure}
    \begin{subfigure}[b]{.32\textwidth}
        \centering
        \includegraphics[trim=0 0 0 200, clip, width=\textwidth]{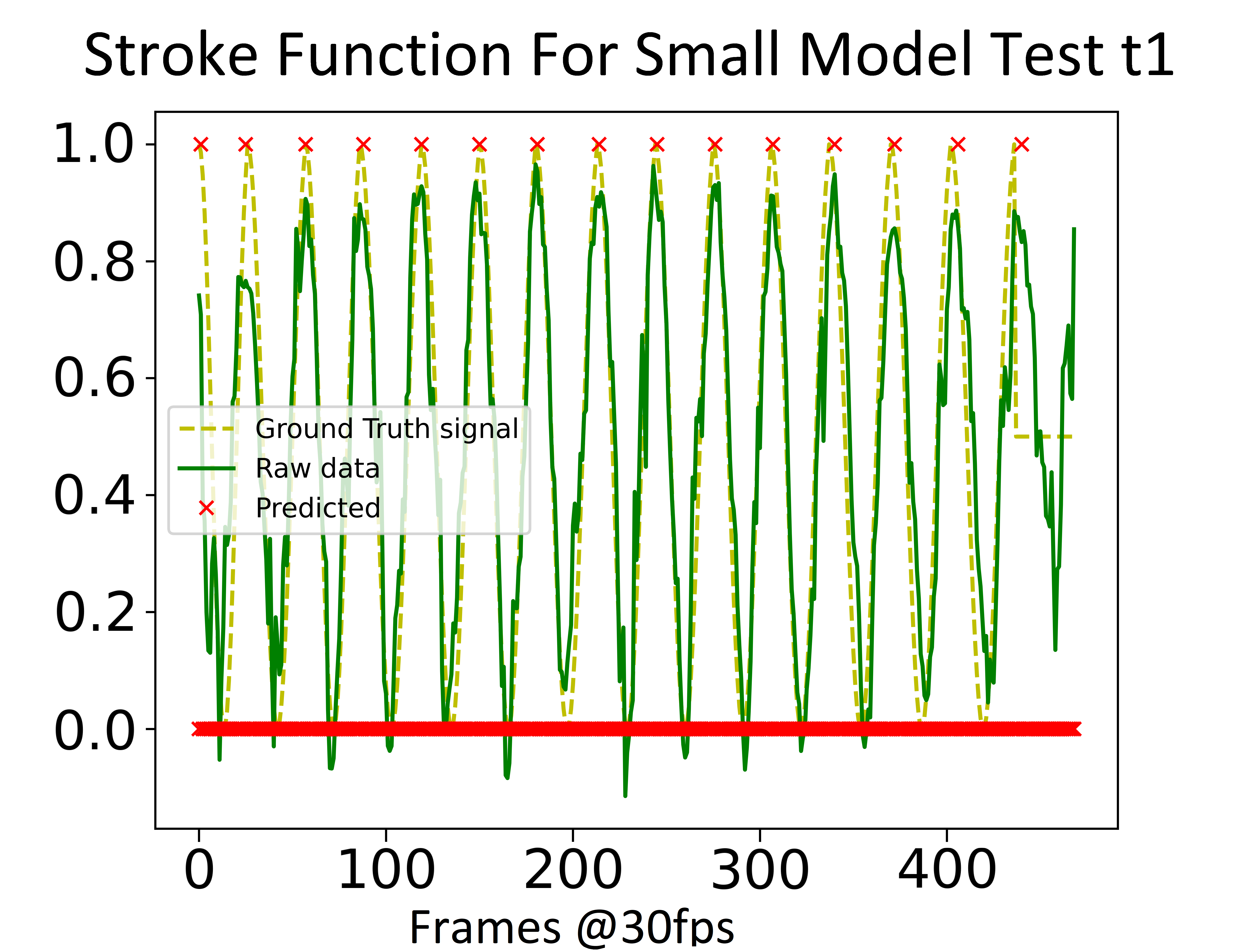}
        \caption{Small model}
        \label{fig:t1_small}
    \end{subfigure}
    \begin{subfigure}[b]{.32\textwidth}
        \centering
        \includegraphics[trim=0 0 0 200, clip, width=\textwidth]{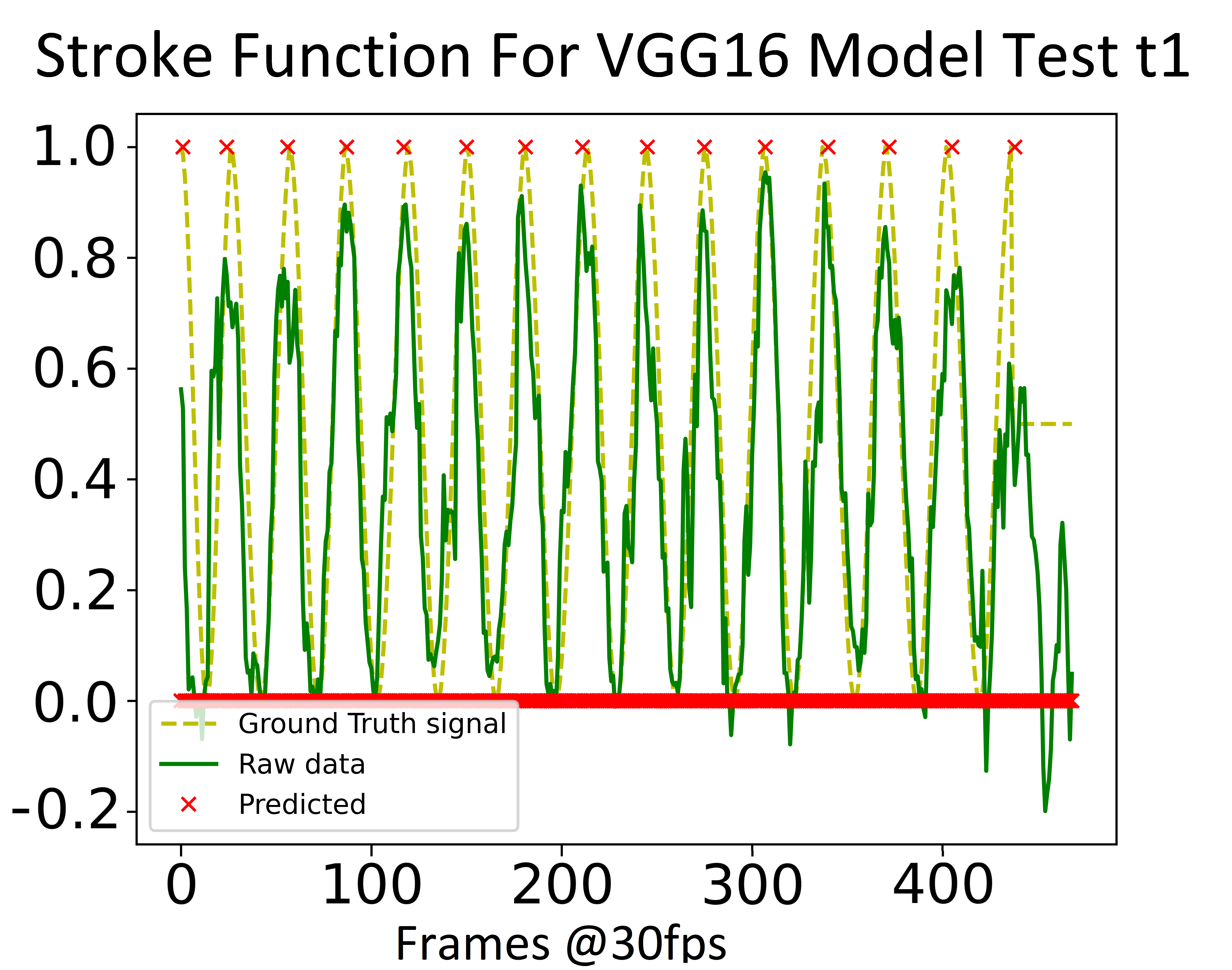}
        \caption{VGG16-based model}
        \label{fig:t1_vgg}
    \end{subfigure}
        \caption{Raw model outputs (green), Butterworth-smoothed outputs (yellow), and top of a stroke (red x) for the tree models.}
        \label{fig:compare_stroke_arch_output}
\end{figure*}

\subsection{Swimmer Tracking} \label{sec:trackExp}
To test swimmer tracking, ground truth data that encodes the trajectory of swimmers, annotated frame by frame, and assigned unique ID numbers, are required. We created such ground truth tracking data from the annotations described in Section~\ref{sec:dataset}. Specifically, the detection annotations were linearly interpolated through all frames using the swimmer lane number provided in the dataset. This created 32 sequences to be used for testing the tracker. Of the 32 sequences, 6 contained frames that were not seen by the detection model, while the other 26 sequences were used in training the detection model. Although the Kalman-based tracker does not require training, we felt that the detector's performance (and therefore the tracker's performance as well) would be quite different on these two sets of sequences, so we report two tracking scores: ``training'' (on the 26 sequences seen by the detector) and ``test'' (on the 6 sequences not seen by the detector). Tracking performance was measured using the MOT metrics dev kit~\cite{MOT_devKit}, with a variety of tracking metrics whose definitions can be found in~\cite{MOT_devKit, leal2015motchallenge, ristani2016performance}. 


\begin{table}[t]
    \centering
    \begin{tabular}{c|c|c|c|c|c}
        Sequences & IDF1 & IDP & IDR & MOTA & MOTP\\
        \hline \hline
        Training & 82.13 & 85.07 & 79.38 & 89.34 & 74.51\\
        \hline
        Test & 39.21 & 42.25 & 36.59 & 11.21 & 61.71\\
    \end{tabular}
    \caption{Tracking results based on metrics from~\cite{MOT_devKit,ristani2016performance}.}
    \label{tab:swim_mot_res_IDF1}
\end{table}


\begin{table} [t]
    \centering
    \begin{tabular}{c|c|c|c|c}
        Sequences & GT & MT & PT & ML\\
        \hline \hline
        Training & 199 & 164 & 35 & 0\\
        \hline
        Test & 48 & 9 & 31 & 8\\
    \end{tabular}
    \caption{Tracking results based on metrics from~\cite{MOT_devKit,ristani2016performance}.}
    \label{tab:swim_mot_res_tracking}
\end{table}

The results are shown in Tables~\ref{tab:swim_mot_res_IDF1}-\ref{tab:swim_mot_res_tracking}. As expected, on the sequences the detector has seen before (``training'') the tracking accuracy is higher across all metrics compared to the sequences not seen by the detector (``test''). A closer examination of the ``test'' results showed that of the 48 trajectories present in this set, 40 were tracked to some degree. The tracker performed much better on the portion of the footage that contained the ``swimming'' class, while it had problems with the footage containing other classes. This could be because other classes were less represented in detector training, so the detector's accuracy was lower here, which negatively affected the tracker. Another issue for tracking failures was occlusions, which may occur for a variety of reasons such as water refraction, opaque water, the race venue, camera field of view, and people on the side of the pool. 

\subsection{Stroke Rate Estimation} \label{sec:strokeRateExp}
Stroke rate estimation was performed on the 346 sub-videos, created from a set of videos different from those used for detection, using the tracking system explained in Section~\ref{sec:tracking}. This data was split into training, validation, and test sets of roughly 80\%, 10\%, and 10\% respectively, of the total amount of data, and three models (Victor, Small, and VGG16) were trained on the test data, as described in Section~\ref{sec:action}. Fig.~\ref{fig:compare_stroke_arch_output} shows samples of the raw model output (green), Butterworth-smoothed output (yellow), and the top of a stroke (red x).  

\begin{table}[t]
    \centering
    \begin{tabular}{c|c|c|c|c}
        Model & F1-score & ASD & SDSTD & $\Delta$\\
        \hline \hline
        Victor & 0.866 & 2.05 & 3.87 & 0.157$^*$\\
        \hline
        Small & 0.859 & 2.18 & 4.75 & 0.149$^*$\\
        \hline
        VGG16 & 0.794 & 2.64 & 4.42 & 0.178$^*$\\
    \end{tabular}
    \caption{Stroke position estimation results. $^*$In the version of the paper on IEEEXplore, the values of $\Delta$ are incorrect; these are the correct values.}
    \label{tab:stroke_detection_res}
\end{table}


To quantify the accuracy, we used metrics defined in~\cite{victor}, and the results are shown in
Table~\ref{tab:stroke_detection_res}. For the F1-score, a predicted stroke peak is considered correct if it is within three frames of a ground-truth stroke peak. The average stroke distance (ASD, in frames) and stroke distance standard deviation (SDSTD, in frames) are measures of the absolute distance from a predicted stroke peak to the nearest ground-truth stroke peak. 
$\Delta$ is the average difference between the predicted s-values and the ground truth. 
All these metrics measure stroke \textit{position} accuracy. Stroke \textit{rate} is the inverse of the time difference between neighboring stroke peaks.

As seen in Table~\ref{tab:stroke_detection_res}, Victor and Small models produce more accurate results than the VGG16-based model. 
The reason for the somewhat weaker performance of the VGG16-based model is likely that part of the backbone was frozen during training, and may therefore not be well-tuned for extracting features necessary for this particular task.
All three models achieve reasonable accuracy and can identify a stroke peak to within 2-3 frames, on average, from its ground-truth position. However, the standard deviation of these predictions is around 4 frames, which should be reduced for highly accurate stroke rate estimation. 

\section{Conclusion} \label{sec:conclusion}
In this work, we presented a swimming analytics system based on unconstrained overhead race video (ORV), as well as a dataset used for training and testing various modules in the system: swimmer detection, classification, tracking, and stroke rate estimation. The proposed system provides a solid basis to build upon and for developing methods for extracting additional analytics from swimming ORV.

\bibliographystyle{IEEEbib}
\bibliography{icme2021template}

\end{document}